\documentclass[prd,twocolumn,twoside,preprintnumbers,nofootinbib,superscriptaddress]{revtex4-1}
\usepackage{physics}
\usepackage{amsmath}
\usepackage{amssymb}
\usepackage{hyperref}
\usepackage{xspace}
\usepackage{slashed}
\hypersetup{
    colorlinks=true,       
    linkcolor=blue,          
    citecolor=blue,        
    filecolor=blue,      
    urlcolor=blue           
}
\usepackage{xcolor}
\usepackage{color}
\usepackage{graphicx}  %
\usepackage{bm}  %
\usepackage{amsmath}   %
\usepackage{graphics}
\usepackage{color}
\usepackage{colortbl}
\usepackage{ulem}

\newcommand{\dslash}[1]{#1 \llap{/\kern-0.5pt}}
\newcommand{\Dslash}[1]{#1 \llap{/\kern+1.5pt}}
\newcommand{\DDslash}[1]{#1 \llap{/\kern+2.3pt}}
\newcommand{\dslashh}[1]{#1 \llap{/\kern+1pt}}

\newcommand{\bea}{\begin{eqnarray}}
\newcommand{\eea}{\end{eqnarray}}
\newcommand{\be}{\begin{equation}}
\newcommand{\ee}{\end{equation}}
\newcommand{\bma}{\begin{pmatrix}}
\newcommand{\ema}{\end{pmatrix}}

\newcommand{ \mysmall}[1]{\scriptscriptstyle #1} 

\allowdisplaybreaks[1]

\begin{document}

\title{Hunting for CP-violating axionlike particle interactions}

\author{Luca Di Luzio}
\affiliation{Dipartimento di Fisica e Astronomia `G. Galilei', Universit\`a di Padova, Italy}
\affiliation{Istituto Nazionale Fisica Nucleare, Sezione di Padova, I--35131 Padova, Italy}
\affiliation{DESY, Notkestra\ss e 85, D-22607 Hamburg, Germany}

\author{Ramona Gr\"{o}ber}
\affiliation{Dipartimento di Fisica e Astronomia `G. Galilei', Universit\`a di Padova, Italy}
\affiliation{Istituto Nazionale Fisica Nucleare, Sezione di Padova, I--35131 Padova, Italy}

\author{Paride Paradisi}
\affiliation{Dipartimento di Fisica e Astronomia `G. Galilei', Universit\`a di Padova, Italy}
\affiliation{Istituto Nazionale Fisica Nucleare, Sezione di Padova, I--35131 Padova, Italy}

\begin{abstract}
The impact of axion-like particles (ALPs) on the search of permanent electric dipole moments (EDMs) of molecules, atoms, nuclei and 
nucleons is systematically investigated. Assuming the most general ALP effective field theory containing operators up to dimension-5, 
we evaluate the leading short-distance effects to the EDMs at two-loop order. 
The high sensitivity of EDMs to CP-violating ALP interactions is emphasised exploiting both the current and projected experimental sensitivities. 
\end{abstract}

\maketitle

\section{Introduction} 
The lack for heavy new physics (NP) at the LHC, mostly motivated by the weak-scale hierarchy problem, has triggered a shift of paradigm 
towards alternative scenarios, with new light mediators, that are receiving increasing attention both theoretically and experimentally.
NP scenarios with light pseudoscalar bosons, referred to as axion-like-particles (ALPs)~\cite{Jaeckel:2010ni} are prominent examples. 
The lightness of ALPs can be naturally explained if they are identified with the pseudo-Nambu-Goldstone bosons of an approximate global symmetry.
Interestingly, ALPs can be invoked to address a number of fundamental open questions in particle physics such as the strong CP problem~\cite{Peccei:1977hh}, 
the origin of dark matter~\cite{Preskill:1982cy}, as well the flavor~\cite{Wilczek:1982rv}  and hierarchy~\cite{Graham:2015cka} problems. 
Furthermore, various anomalies can be solved by the ALPs such as the longstanding discrepancy of the anomalous magnetic moment 
of the muon~\cite{Chang:2000ii,Marciano:2016yhf}, the excess in excited Beryllium decays ${^8\text{Be}}^*\to {^8\text{Be}}+e^+ e^-$ \cite{Krasznahorkay:2015iga,Feng:2016ysn,Ellwanger:2016wfe} and that of electronic recoil events with an energy of $\mathcal{O}(\rm keV)$ 
observed by the XENON1T collaboration~\cite{Aprile:2020tmw,Takahashi:2020bpq}.

Since the relation between ALP mass and couplings depends on the specific ultraviolet (UV) completion, it is customary to take a model-independent 
approach where ALPs are treated as a generalization of the QCD axion, with mass and couplings being free parameters to be probed experimentally. 
In this framework, ALP interactions with Standard Model (SM) fermions and gauge bosons are described via an effective Lagrangian built with operators 
up to dimension-5~\cite{Georgi:1986df}. This approach still captures general features of a broad class of models. 

For ALP masses below the MeV scale, a vast experimental program, intertwined with cosmology and astrophysics, is currently ongoing~\cite{Jaeckel:2010ni}. 
That ranges from ``wave-like'' approaches to ALP searches in the sub-eV region (such as haloscopes, helioscopes and optical/EM setups), to beam-dump 
experiments stretching up to the GeV scale \cite{Dobrich:2019dxc}. Collider experiments have also probed ALP masses ranging from the GeV scale up to the 
electroweak scale, through searches of ALPs associated production with photons, jets and electroweak gauge bosons \cite{Jaeckel:2015jla}. 
Searches for the exotic, on-shell Higgs and $Z$ decays into ALPs were also shown to probe regions of the parameter space previously unconstrained~\cite{Bauer:2017ris}.
Other low-energy observables which are extremely sensitive to ALPs are flavor-changing neutral-currents (FCNC) processes both in the quark~\cite{Batell:2009jf} and lepton sectors~\cite{Bauer:2019gfk}. Indeed, since there is no fundamental reason for the ALP interactions to respect the SM flavor group, ALPs can induce FCNC already at tree level.

Rather surprisingly, CP-violating (CPV) signatures of ALPs received so far much less attention~\cite{Marciano:2016yhf}.\footnote{CPV interactions have been investigated in the context of long-range forces for the QCD axion~\cite{Moody:1984ba} 
and for scalar-pseudoscalar ALP couplings to nucleons and electrons~\cite{Stadnik:2017hpa}.}
The CP symmetry is violated as long as ALP couplings to photons entail both $\phi F\tilde F$ and $\phi F F$ interactions
(where $\phi$ is the ALP, $F$ the QED field strength tensor and $\tilde F$ its dual) and/or if ALP couplings to fermions ($f$) include 
both $\phi \bar f \gamma_5 f$ and $\phi \bar f f$ interactions. As we will see, these conditions require that the global shift
symmetry (responsible for the lightness of the ALP) as well as CP are broken in the UV sector and such a symmetry breaking 
is eventually communicated to the infrared dynamics by some mediators.
The required UV dynamics can arise quite naturally in strongly-coupled theories, 
and in fact an explicit realization is provided by the SM itself. 
Consider for definiteness the effective interactions of the neutral pion field $\pi^0$ below the GeV scale (see e.g.~\cite{Choi:1990cn}). 
The role of the ALP is played by $\pi^0$, while quark masses are responsible for the breaking of the shift symmetry and the source of 
CP violation is the QCD $\theta$ term. The mediators from the strong sector to the $\pi^0$ are instead the electromagnetic (EM) interactions. 
CP-even pion interactions contain the terms $A_1 \frac{\pi^0}{f_\pi} F \tilde F + A_2 \frac{\partial_\mu \pi^0}{f_\pi} \bar e \gamma^\mu \gamma_5 e$ 
where $A_1 = \frac{\alpha}{4\pi}$ is the Wess-Zumino-Witten term and $A_2$ is generated radiatively from $A_1$ via EM 
interactions so that $A_2 \sim (\frac{\alpha}{4\pi})^2$.
CP-odd pion interactions $C_1 \frac{\pi^0}{f_\pi} F F + C_2 m_e \frac{\pi^0}{f_\pi} \bar e e$ are instead sourced by the QCD $\theta$ 
term $C_1 \sim \theta$ while $C_2$ is generated radiatively from $C_1$ via EM and therefore $C_2 \sim \theta \frac{\alpha}{4\pi}$.
A strongly coupled dynamics at the scale $\Lambda \gtrsim 1$ TeV that resembles the pion dynamics of the SM can hence be conceived in analogy.

Other scenarios providing a strong motivation for the study of a CPV ALP are relaxion models \cite{Graham:2015cka}
which propose a new solution to 
the weak-scale hierarchy problem by introducing an ALP field, the relaxion, which scans the Higgs boson mass in the early universe 
from an initial value comparable to the cutoff scale $M \gg 1$ TeV to the final value of $\mathcal{O}(v)$ with $v=246$ GeV. 
The simultaneous presence of the relaxion-Higgs mixing and the relaxion-photon/gluon couplings violates CP.

From the experimental side, there is an extraordinary ongoing program aiming at improving the current limits of permanent electric dipole moments (EDMs) 
of molecules, atoms, nuclei and nucleons by many orders of magnitude \cite{Chupp:2017rkp}. 
In the light of the above considerations, we believe that a still-missing comprehensive exploration of the phenomenological implications of a CPV ALP is mandatory. The main motivation of the present work is to fill this gap.

The paper is structured as follows. After introducing in Section \ref{sec:alpscpvcoupl} the ALP effective field theory (EFT) containing operators up to dimension-5, in 
Section \ref{sec:alpeft} we evaluate the leading short-distance effects to the EDMs up to two-loop order. This will enable us to highlight in Section \ref{sec:alpedms} the high sensitivity of EDMs to CPV ALP interactions once the current 
and projected experimental sensitivities on EDMs are exploited. We conclude in Section \ref{sec:concl} with an outlook 
for possible generalizations of our work. 

\section{ALPs with CP-violating couplings}
\label{sec:alpscpvcoupl}
The most general $SU(3)_c \times U(1)_{\rm em}$ invariant dimension-5 effective Lagrangian accounting for CPV ALP interactions 
with photons, gluons and SM fermions, reads
%
%
%
\begin{align} 
\label{eq:Lphi}
\mathcal{L}_{\phi} & =  
e^2 \frac{\tilde C_{\gamma}}{\Lambda} \phi F \tilde F + g^2_s \frac{\tilde C_{g}}{\Lambda} \phi G \tilde G  + i \frac{v}{\Lambda} y^{ij}_P \phi \bar f_i \gamma_5 f_j 
\nonumber\\
& + e^2 \frac{C_{\gamma}}{\Lambda} \phi F F + g^2_s \frac{C_{g}}{\Lambda}  \phi G G  + \frac{v}{\Lambda} y^{ij}_S \phi \bar f_i f_j
\end{align}
where $f \in (e,u,d)$ denotes SM fermions in the mass basis, $i,j$ are flavor indices and, by construction, the matrices $y_{S}$ and $y_{P}$ are hermitian. 
$F$ and $G$ stand for the QED and QCD field strenght tensors, respectively, while $\tilde F$ and $\tilde G$ are their duals. 
Note that the interactions in the first line of eq.~(\ref{eq:Lphi}) can be understood to be invariant under the $\phi$ shift symmetry, as long as 
non-perturbative effects related to boundary terms can be neglected. 
In fact, both $F \tilde F$ and $G \tilde G$ are total derivatives, while pseudoscalar interactions can be written in a shift-symmetric way through the dimension-5 operator $\frac{\partial_\mu \phi}{\Lambda} \bar f \gamma^\mu \gamma_5 f$ upon integrating by parts and applying the equations of motion. This justifies their normalization factor $\frac{v}{\Lambda}$. Instead, the interactions in the second line of eq.~(\ref{eq:Lphi}) break explicitly the shift symmetry. In particular, scalar interactions can be written, in the unbroken SM phase, in terms of the dimension-5 operator $\phi H\bar f_{L(R)} f_{R(L)}$ thus justifying the normalization factor $\frac{v}{\Lambda}$. We remark that $(\tilde{C}_\gamma, \tilde{C}_g, y_P)$ and $(C_\gamma, C_g, y_S)$ could have very different size as they stem from the shift-symmetry invariant and breaking sectors, respectively.
Finally, ALP-Higgs mixing, induced by scalar potential operators of the type $|H|^2 \sin(\phi/\Lambda)$, can be straightforwardly taken into account in the EDM formulae below~\cite{paper_long}.

%
%

Hereafter, we assume that $m_\phi \gtrsim \rm few~$GeV so that QCD can be treated perturbatively.
Moreover, although we take $\Lambda\gtrsim 1$ TeV, we focus only on electromagnetic and strong interactions as weak interactions 
play a subleading role in our analysis.

The effective Lagrangian $\mathcal{L}_{\phi}$ at the scale $\Lambda$ is renormalized at lower energies by QED and QCD interactions.
Although the full one-loop anomalous dimension matrix for the dimension-5 operators of our ALP EFT will be presented elsewhere~\cite{paper_long},
in the following we discuss the most relevant effects for our analysis. In the leading logarithmic approximation, 
the solution of the renormalization-group equations for the leptonic (pseudo)scalar couplings $y^{\ell\ell}_{S,P}$ gives
\begin{align}
\label{eq:y_l_RGE}
y^{\ell\ell}_S & \simeq y^{\ell\ell}_S(\Lambda) + \frac{6\alpha}{\pi} \frac{m_\ell}{v} e^2 C_\gamma \log\frac{\Lambda}{\mu}\,,
\\
y^{\ell\ell}_P & \simeq y^{\ell\ell}_P(\Lambda) - \frac{6\alpha}{\pi} \frac{m_\ell}{v} e^2 {\tilde C}_\gamma \log\frac{\Lambda}{\mu}\,,
\end{align}
where $\mu$ is the renormalization scale. Instead, in the quark sector, we obtain
\begin{align}
\!\! y^{qq}_S \simeq & \,\, y^{qq}_S(\Lambda) \!+\! 
\frac{m_q}{v} \! \left( \frac{6\alpha}{\pi} Q^2_q e^2 C_\gamma \!+\! \frac{8\alpha_s}{\pi} g_s^2 C_g \right) \!
\log\frac{\Lambda}{\mu}\,,
\\
\! y^{qq}_P \simeq & \,\, y^{qq}_P(\Lambda) \!-\! 
\frac{m_q}{v} \! \left( \frac{6\alpha}{\pi} Q^2_q e^2 {\tilde C}_\gamma \!+\! \frac{8\alpha_s}{\pi} g_s^2 {\tilde C}_g \right)\!
\log\frac{\Lambda}{\mu}\,.
\label{eq:y_q_RGE}
\end{align}
Note that QED and QCD loop-corrections in eqs.~(\ref{eq:y_l_RGE})-(\ref{eq:y_q_RGE}) are significantly larger 
than the expectations from naive dimensional analysis. 

\begin{figure*}
\includegraphics[width=1.0\linewidth]{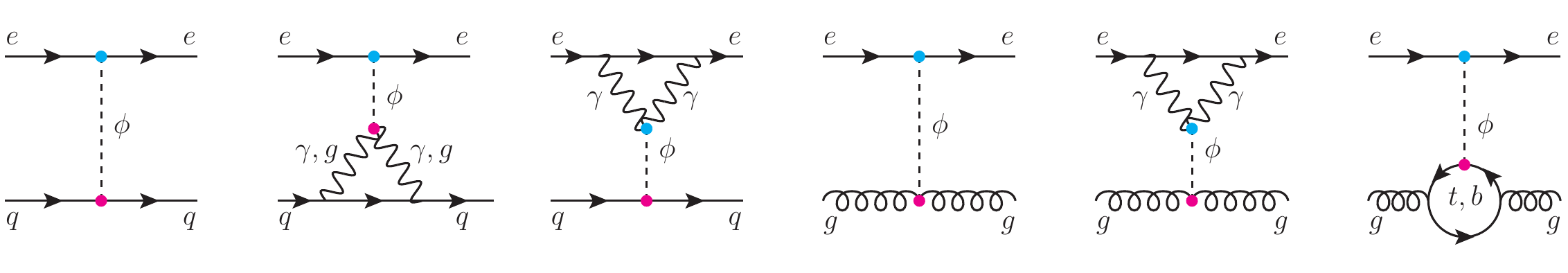}
\caption{Leading contributions to the semi-leptonic nucleon-electron operators.
The combination of light-blue and purple blobs refer to CPV effective interaction vertices.}
\label{fig:4f}
\end{figure*}

Since in eq.~(\ref{eq:Lphi}) we factor out the gauge couplings $e^2$ and $g^2_s$, the coefficients $C_{\gamma,g}$ and $\tilde C_{\gamma,g}$ turn out to be 
scale invariant at one-loop order. Yet, top and bottom contributions are taken into account by the QCD trace anomaly in the gluon-gluon-ALP vertex after they 
have been integrated out (see fig.~\ref{fig:4f}). The resulting effect is
\begin{align}
g_s^2 C_g & \simeq g_s^2 C_g(\Lambda) + \frac{\alpha_s}{12\pi} \sum_{q=t,b} \frac{v y^{qq}_S}{m_q} f_g(x_q) \,,
\\
g_s^2 \tilde C_g & \simeq g_s^2 \tilde C_g(\Lambda) - \frac{\alpha_s}{8\pi} \sum_{q=t,b} \frac{v y^{qq}_P}{m_q} \tilde{f}_g(x_q)\,,
\end{align}
where $x_q = m^2_\phi / m^2_q$ and $f_g(0)=\tilde{f}_g(0)=1$~\cite{paper_long},
in agreement with the Higgs low-energy theorem \cite{Shifman:1978zn}. 

Since the operators $XX$ and $X \tilde X$ ($X= F,G$), as well as scalar and pseudoscalar operators, have opposite CP transformation 
properties, it is clear that $\mathcal{L}_{\phi}$ violates CP regardless of the CP-even or CP-odd nature of $\phi$.

CP violating phenomena can be conveniently described in terms of Jarlskog invariants, i.e.~rephasing-invariant parameters 
which provide a measure of CP violation~\cite{Jarlskog:1985ht}. 
In particular, the full set of Jarlskog invariants of our ALP EFT reads 
%
\begin{align} 
\label{eq:Jarlskog}
C_{a}\tilde C_{b}, ~~~~ y^{ii}_S \, {\tilde C_{a}}, ~~~~y^{ii}_P \, C_{a}, ~~~~ y^{ii}_S \, y^{jj}_P, ~~~~ y^{ik}_S \, y^{kk}_{\rm \mysmall SM} \, y^{ki}_P\,,
\end{align}
where 
$a,b = \gamma,g$ 
and $y^{kk}_{\rm \mysmall SM}$ denotes a SM Yukawa coupling in the diagonal basis. 
Notice that only the last invariant of eq.~(\ref{eq:Jarlskog}) is sensitive to flavor-violating effects. 
Moreover, as we will see, at the two-loop level all the above invariants will be generated.

\section{Effective Lagrangian for EDMs}
\label{sec:alpeft}
The leading low-energy CPV Lagrangian relevant for EDMs of molecules, atoms, nuclei and nucleons reads~\cite{Pospelov:2005pr}
\begin{align}
\!\mathcal{L}_{\rm CPV}  & =
\sum_{i,j=u,d,e} \!\!\! C_{ij} (\bar f_i f_i) (\bar f_j i \gamma_5 f_j) 
+ \alpha_s C_{Ge} \, GG \, \bar e i\gamma_5e 
\nonumber\\
& + \alpha_s C_{\tilde Ge} \, G\tilde G \, \bar e e 
- \frac{i}{2} \! \sum_{i=u,d,e} d_i \bar f_i (F \!\cdot\! \sigma) \gamma_5 f_i 
\nonumber\\
&
- \frac{i}{2} \! \sum_{i=u,d} g_s d^C_i \bar f_i (G \!\cdot\! \sigma) \gamma_5 f_i
+ \frac{d_G}{3} f^{abc} G^a \tilde G^b G^c 
 \,, 
\label{eq:CPV_lagrangian}
\end{align}
where we omitted color-octet 4-quark operators (as they are induced only at one-loop level in the ALP framework)
and the dim-4 $G\tilde G$ operator. 
The latter is assumed to be absent thanks to a UV mechanism solving the strong CP problem.
Within our EFT, $C_{ij}$, $C_{Ge}$ and $C_{\tilde Ge}$ are generated by the Feynman diagrams 
of fig.~\ref{fig:4f} and read
\begin{align}
\label{eq:Cij}
C_{ij} \simeq  \frac{v^2}{\Lambda^2} \frac{y^{ii}_S y^{jj}_P}{m^2_\phi}\,, \qquad
C_{Ge} = \frac{4\pi}{m^2_\phi} \frac{v}{\Lambda^2} C_g y^{ee}_P \,,
\end{align}
while $C_{\tilde Ge} \leftrightarrow C_{Ge}$ via the replacement $C_g y^{ee}_P \leftrightarrow\tilde C_g y^{ee}_S$.

\begin{figure}[t]
\includegraphics[width=0.92\linewidth]{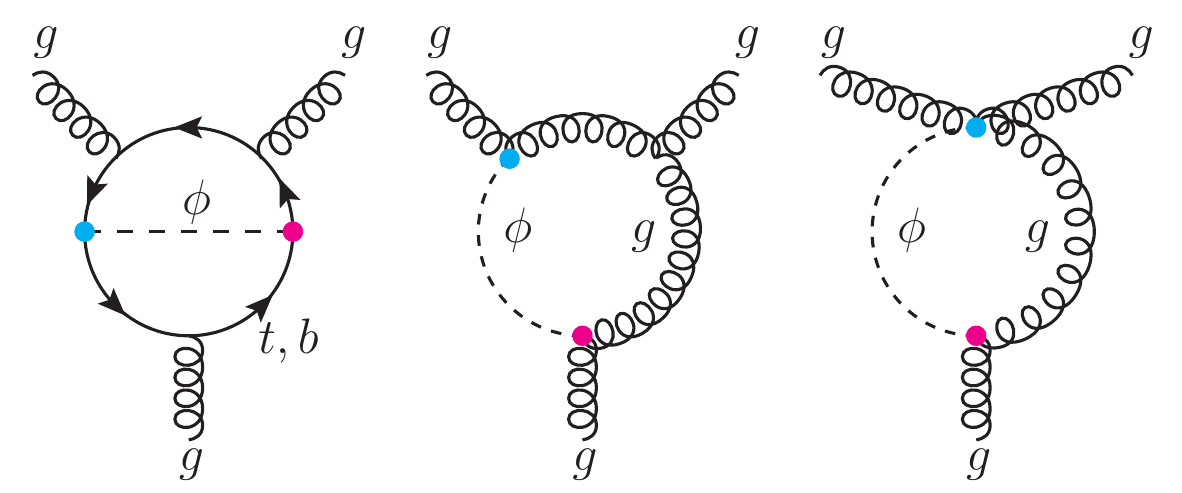}
\caption{Leading contributions to the Weinberg operator.}
\label{fig:Weinberg}
\end{figure}

The last term of eq.~(\ref{eq:CPV_lagrangian}) refers to the Weinberg operator which is generated by the representative diagrams  
shown in fig.~\ref{fig:Weinberg}. The related Wilson coefficient $d_G$ reads
%
%
\begin{align}
\!\!\!\! d_G & \simeq  \! \frac{g_s \alpha_s}{(4\pi)^3} \!\! \sum_{i=t,b} \! \frac{v^2}{\Lambda^2} \frac{y^{ii}_S y^{ii}_P}{4 m^2_{i}} h(x_q)
+ \frac{3g_s}{\pi^2} \frac{g_s^4 C_g \tilde C_g}{\Lambda^2} \log\!\frac{\Lambda}{m_\phi} 
\label{eq:weinberg}
\end{align}
%
where the first term refers to the two-loop diagram and $h(0)=1$~\cite{paper_long}.
Instead, the second term of eq.~(\ref{eq:weinberg}) arises from the one-loop diagrams of fig.~\ref{fig:Weinberg} 
and enjoys a very large enhancement factor with respect to the naive dimensional analysis expectation. 
As a result, we anticipate that $d_G$ will provide the by far dominant effects to EDMs as induced by $C_g \tilde C_g$.

\begin{figure}[t]
\quad \includegraphics[width=0.82\linewidth]{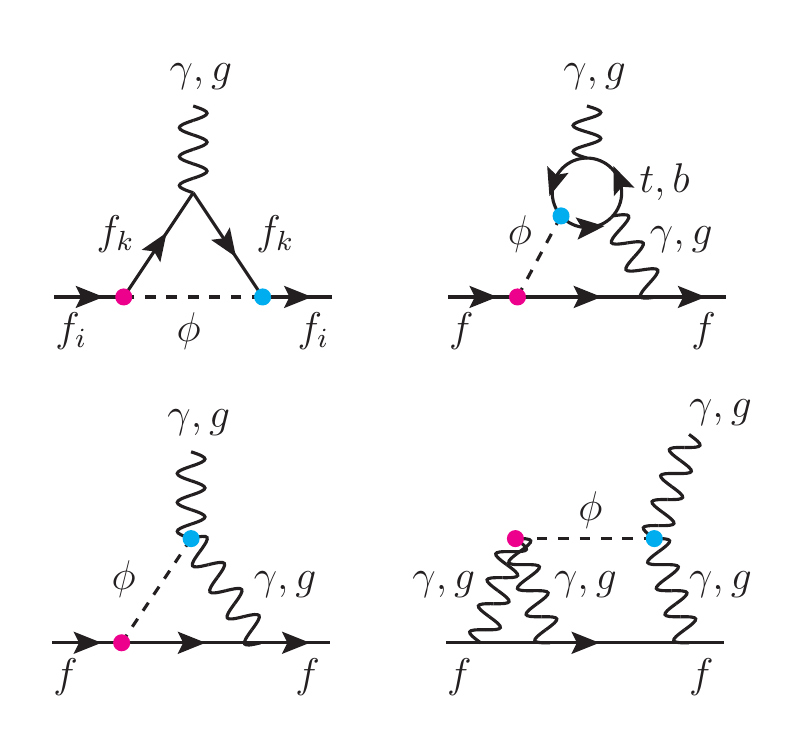}
\caption{Leading contributions to the fermionic (C)EDMs.}
\label{fig:dipole}
\end{figure}

Finally, we analyse the fermionic (C)EDMs induced by ALP interactions. The leading contributions stem from the Feynman 
diagrams reported in fig.~\ref{fig:dipole} and read
\begin{align}
\frac{d_i}{e} \simeq &
- \sum_{k}
\frac{Q_k}{16 \pi^2}\frac{m_k}{m^2_{\phi}}
\frac{v^2}{\Lambda^2}
\Re (y^{ik}_S y^{ki}_P) \, \ell(x_k)
\nonumber\\
&- \sum_{k} \frac{ N^k_c \alpha Q_i Q^2_k}{8\pi^3 m_k}  
\frac{v^2}{\Lambda^2}
\left( 
y^{ii}_P y^{kk}_S f(x_{k}) + y^{ii}_S y^{kk}_P g(x_{k})  
\right)
\nonumber\\
&-\frac{Q_i}{2\pi^2} \frac{v}{\Lambda^2} e^2 (y^{ii}_S {\tilde C}_\gamma - C_\gamma y^{ii}_P) \log\frac{\Lambda}{m_\phi}
\nonumber\\
&-\frac{3 \alpha Q^3_i}{\pi^3} \frac{m_i}{\Lambda^2} e^4 C_\gamma {\tilde C}_\gamma \log^2\frac{\Lambda}{m_\phi}
\nonumber\\
& - \delta_{qi} \frac{2 \alpha_s Q_i}{\pi^3} \frac{m_i}{\Lambda^2} e^2 g^2_s 
( C_\gamma {\tilde C}_g + C_g {\tilde C}_\gamma  )
\log^2\frac{\Lambda}{m_\phi} \, ,
\end{align}
in the EDMs case (where $i=e,u,d$ and $q=u,d$) and
\begin{align}
\label{eq:dCi}
d^C_i \simeq &
- \sum_{k}
\frac{1}{16 \pi^2}\frac{m_k}{m^2_{\phi}} \frac{v^2}{\Lambda^2} \Re (y^{ik}_S y^{ki}_P) \, \ell(x_k)
\nonumber\\
&- \sum_{k} \frac{\alpha_s}{16\pi^3 m_{k}} \frac{v^2}{\Lambda^2}  
\left( 
y^{ii}_P y^{kk}_S f(x_{k}) + y^{ii}_S y^{kk}_P g(x_{k})  
\right)
\nonumber\\
&-\frac{1}{2\pi^2} \frac{v}{\Lambda^2} g_s^2 (y^{ii}_S {\tilde C}_g - C_g y^{ii}_P) \log\frac{\Lambda}{m_\phi}
\nonumber\\
&-\frac{4 \alpha_s}{\pi^3} \frac{m_i}{\Lambda^2} g_s^4 C_g {\tilde C}_g \log^2\frac{\Lambda}{m_\phi}
\nonumber\\
&-\frac{3 \alpha Q^2_i}{2\pi^3} \frac{m_i}{\Lambda^2} e^2 g_s^2 
( C_\gamma {\tilde C}_g + C_g {\tilde C}_\gamma )
\log^2\frac{\Lambda}{m_\phi} \, ,
\end{align}
for the CEDMs (where $i=u,d$). 
The loop functions are 
$\ell(x) = (3 \!-\! 4x \!+\! x^2 \!+\! 2 \log x) / (1-x)^3$ 
and, in the asymptotic limit $x \gg 1$,
$f(x) \approx (6 \log x + 13)/18$ and $g(x) \approx (\log x +2)/2$, 
where $x_{k}= m^2_{k} / m^2_\phi$.
%
While the contributions to the electron EDM stemming from the third and fourth diagrams of fig.~\ref{fig:dipole} were already considered in~\cite{Marciano:2016yhf}, the expressions of quark (C)EDMs are new.
Moreover, we also consider here flavor-violating effects for the first diagram (for flavour-diagonal effects, see~\cite{Chen:2015vqy})
and Barr-Zee two-loop contributions (second diagram). 

Our results are obtained using a hard cutoff as a UV regulator to render loop calculations finite, i.e. the size of the loop diagrams is estimated in terms of 
this hard cutoff and assumes no significant cancellations with finite terms that could change these estimates.

Although eqs.~(\ref{eq:y_l_RGE})-(\ref{eq:dCi}) capture only the leading-order short-distance effects of our ALP model, the bounds of Table~\ref{Tab:ALL} have been obtained taking into account also one-loop QCD running effects (improved with a two-loop running of $\alpha_s$ and quark masses) from $\Lambda = 1$ TeV down to $m_\phi= 5\,$GeV
and the running of $\mathcal{L}_{\rm CPV}$ from $m_\phi$ down to the hadronic scale $\mu_{\rm had} = 1$ GeV~\cite{Degrassi:2005zd}.

Constraints on $d_e$, as well as on the coefficients $C_{ij}$ and $C_{Ge}$ in eq.~(\ref{eq:CPV_lagrangian}), are set by using the polar 
molecule ThO. The electron spin-precession frequency receives contributions from both $d_e$ and CP-odd electron--nucleon ($N$) 
interactions $\mathcal{L} \supset - \frac{G_F}{\sqrt{2}} C_S \bar N N \bar e i\gamma_5e$~\cite{Dekens:2018bci}
\begin{align}
\omega_{\text{ThO}} = 1.2 \,\mathrm{mrad}/\mathrm{s}
\left(\frac{d_e}{10^{-29}\,e\,\mathrm{cm}}\right) +1.8 \,\mathrm{mrad}/\mathrm{s}\left(\frac{C_S}{10^{-9}}\right)
\nonumber
\end{align}
with a theoretical error of few percent and the experimental limit $\omega_{\text{ThO}} < 1.3\,\mathrm{mrad}/\mathrm{s}$ 
($90\%$ C.L.)~\cite{Andreev:2018ayy}. The coefficient $C_S$ is related to $C_{ij}$ and $C_{Ge}$ as 
$C_S / v^2 \simeq -17 (C_{ue} + C_{de}) + 4.7 \, \text{GeV} \, C_{Ge}$.
The neutron EDM is induced by quark (C)EDMs, the Weinberg operator and 4-quark operators~\cite{Hisano:2012cc,Cirigliano:2019vfc,Bertolini:2019out} 
\begin{align}
d_n & = 0.784(28)\,d_u - 0.204(11)\,d_d 
-0.55(28)\,e\, d^C_u  
\notag\\
&- 1.10(55)\,e\, d^C_d   
+ 50(40)\,{\rm MeV}\,e\,d_{G} \notag\\
&+ 30(20)\,{\rm MeV} 
\,e\,(C_{ud}-C_{du}) \, , 
\end{align}
while the experimental bound is $d_n < 1.8 \cdot 10^{-26}\,e~$cm ($90\%$ C.L.)~\cite{Abel:2020gbr}. 
The EDM of the diamagnetic atom ${}^{199}$Hg gets contributions from both nuclear and leptonic CP-odd interactions~\cite{Dekens:2018bci,Cirigliano:2019vfc} 
\begin{align}
d_{\rm Hg} \simeq 4.0 \cdot 10^{-4} d_{n} - [ 2.8 \, C_S - 2.1 \, C_P ] \, 10^{-22}\,e\,{\rm cm}
\end{align}
with 
$C_P \simeq C^{(0)}_P - C^{(1)}_P$ defined in terms of the Lagrangian 
$\mathcal{L} \supset - \frac{G_F}{\sqrt{2}} \bar N ( C^{(0)}_P + \tau_3 C^{(1)}_P )  i\gamma_5 N \bar e e$.
To set bounds we employ
$C_P/v^2 = 350 (C_{eu} + C_{ed}) + 1.1 \, \text{GeV} \, C_{\tilde Ge}$ and the experimental limit 
$d_{\rm Hg}< 6 \cdot 10^{-30}e~$cm ($90\%$ C.L.)~\cite{Graner:2016ses}.

\section{Probing ALPs with EDMs} 
\label{sec:alpedms}
The sensitivity of physical EDMs to the CPV invariants of eq.~(\ref{eq:Jarlskog}) are reported in Table~\ref{Tab:ALL}, 
where we have taken $m_\phi = 5$ GeV and employed central values for theoretical predictions. While the bounds on $|C_\gamma {\tilde C}_\gamma|$ 
and $|y^{ee}_S {\tilde C}_\gamma - y^{ee}_P C_\gamma|$ were already studied in~\cite{Marciano:2016yhf}, all the other bounds are new. 
As shown in Table~\ref{Tab:ALL}, 4-fermion operators provide the most stringent bounds on several CP invariants. 
Similarly, the Weinberg operator sets tight limits on ALP couplings to the top and bottom quarks as well as to gluons 
which were previously unconstrained. We remark that also flavor-violating contributions to the EDMs are quite effective. 
Indeed, despite the suppression arising from flavor mixing angles --which are otherwise constrained by FCNC processes-- 
(C)EDMs enjoy a chiral enhancement $m_k/m_i$, for $k>i$, which is absent in the case of flavor-conserving interactions. 
\begin{table}[t]
\small
	\centering
	\begin{tabular}{c c c c}
		\toprule
		 CPV invariant &~~~ Bound  &~~~ Observable \\
		\colrule
	 	$|C_\gamma {\tilde C}_\gamma|$   & ~~~ $6.2 \times 10^{-3}$ & ~~~ 
		$\omega_{\rm ThO}(d_e)$   \\
		$|C_g {\tilde C}_g|$   & ~~~ $1.4 \times 10^{-6}$ &~~~ $d_n,d_{\rm Hg}(d_G)$ \\
		$|C_\gamma {\tilde C}_g|$ & ~~~ $0.40$  &~~~ $d_{\rm Hg}(C_P, C_S)$ \\ 
		$|C_g \tilde C_\gamma|$   & ~~ $2.3 \times 10^{-3}$ &~~~ $\omega_{\rm ThO}(C_S)$ \\ 
		$|y^{ee}_S {\tilde C}_\gamma - y^{ee}_P C_\gamma|$   & ~~~ $6.9 \times 10^{-11}$  & ~~~ $\omega_{\rm ThO}(d_e)$  \\	
		$|y^{uu}_S {\tilde C}_g - y^{uu}_P C_g|$ & ~~~ $8.1 \times 10^{-9}$  & ~~~ $d_n, d_{\rm Hg} (d^C_{u})$ \\
		$|y^{dd}_S {\tilde C}_g - y^{dd}_P C_g|$ & ~~~ $6.5 \times 10^{-9}$  & ~~~ $d_n, d_{\rm Hg} (d^C_{d})$ \\
		$|y^{ee}_P C_g|$   & ~~~ $2.1 \times 10^{-11}$  &~~~ $\omega_{\rm ThO}(C_S)$ \\ 
		$|y^{ee}_S \tilde C_g|$   & ~~~ $7.3 \times 10^{-9}$   &~~~ $d_{\rm Hg}(C_P)$ \\ 
		$|y^{uu}_S y^{dd}_P - y^{dd}_S y^{uu}_P|$ & ~~~ $5.6 \times 10^{-9}$  & ~~~ $d_{\rm Hg} (C_{ud} - C_{du})$ \\
		$|y^{ee}_S y^{uu}_P|$, $|y^{ee}_S y^{dd}_P|$  & ~~~ $4.2 \times 10^{-13}$  &~~~ $d_{\rm Hg}(C_P)$ \\
		$|y^{uu}_S y^{ee}_P|$, $|y^{dd}_S y^{ee}_P|$   & ~~~ $2.1 \times 10^{-13}$ &~~~  $\omega_{\rm ThO}(C_S)$ \\		
		$|y^{tt}_S y^{ee}_P|$   & ~~~ $6.8 \times 10^{-9}$  &~~~ $\omega_{\rm ThO}(C_S)$ \\
		$|y^{bb}_S y^{ee}_P|$   & ~~~ $1.7 \times 10^{-10}$  &~~~ $\omega_{\rm ThO}(C_S)$ \\
		$|y^{ee}_S y^{tt}_P|$   & ~~~ $8.8 \times 10^{-9}$  & ~~~  $\omega_{\rm ThO}(d_e)$ \\
		$|y^{ee}_S y^{bb}_P|$   & ~~~ $3.9 \times 10^{-9}$  &~~~ $d_{\rm Hg}(C_P)$ \\
		$|y^{tt}_S y^{tt}_P|$   & ~~~ $0.10$ & ~~~ $d_n,d_{\rm Hg}(d_G)$  \\
		$|y^{bb}_S y^{bb}_P|$   & ~~~ $5.9\times 10^{-5}$ & ~~~ $d_n,d_{\rm Hg}(d_G)$   \\
		$|y^{ee}_S \, y^{ee}_P |$   &~~~  $1.0 \times 10^{-10}$  & ~~~ $\omega_{\rm ThO}(d_e)$  \\	
		$|y^{e\mu}_S \, y^{\mu e}_P |$   &~~~  $2.2 \times 10^{-12}$   & ~~~ $\omega_{\rm ThO}(d_e)$  \\	
		$|y^{e\tau}_S \, y^{\tau e}_P |$   &~~~  $3.1 \times 10^{-12}$  & ~~~ $\omega_{\rm ThO}(d_e)$  \\	
		$|y^{uu}_S \, y^{uu}_P |$   &~~~  $3.9 \times 10^{-8}$  & ~~~ $d_n, d_{\rm Hg} (d^C_{u})$  \\			
		$|y^{uc}_S \, y^{cu}_P |$   &~~~  $3.2 \times 10^{-9}$  & ~~~ $d_n, d_{\rm Hg} (d^C_{u})$  \\	
		$|y^{ut}_S \, y^{tu}_P |$   &~~~  $3.2 \times 10^{-7}$ & ~~~ $d_n, d_{\rm Hg} (d^C_{u})$  \\
		$|y^{dd}_S \, y^{dd}_P |$   &~~~  $4.6 \times 10^{-8}$    & ~~~ $d_n, d_{\rm Hg} (d^C_{d})$  \\			
		$|y^{ds}_S \, y^{sd}_P |$   &~~~  $6.4 \times 10^{-9}$  & ~~~ $d_n, d_{\rm Hg} (d^C_{d})$  \\
		$|y^{db}_S \, y^{bd}_P |$   &~~~  $1.7 \times 10^{-8}$   & ~~~ $d_n, d_{\rm Hg} (d^C_{d})$  \\				
		\hline				
		\botrule
	\end{tabular}\\
	\caption{Bounds on CPV invariants for $\Lambda = 1$ TeV and $m_\phi = 5$ GeV. 
	 In the 3rd column we specify the observable and (in brackets) the leading operator setting the bound.}
	\label{Tab:ALL}
\end{table}
Although the relative importance of the above contributions to the physical EDMs will depend on the relative strength of the microscopic 
parameters $C_{\gamma(g)}$, $\tilde C_{\gamma(g)}$, $y_{S}$ and $y_{P}$ and therefore on the specific ALP model, let us consider as an example 
the case where $y^{ii}_{S,P} \propto \frac{m_i}{v}$, $C_{\gamma(g)}$ and $\tilde C_{\gamma(g)} \propto \frac{1}{16\pi^2}$. In such a setup it turns out 
that 4-fermion operators are the by far best probes of CP violation followed by the electron EDM and the Weinberg operator which have comparable sensitivities, 
as it can be easily checked from Table~\ref{Tab:ALL}. 
We finally remark that the electron EDM bound on CPV combinations studied here are always much stronger than the corresponding limits on CP-conserving 
combinations stemming from the anomalous magnetic moments of the electron, unless CPV phases are smaller than about $10^{-4}$~\cite{Giudice:2012ms}.

The expected sensitivities of future EDM experiments will greatly improve the current resolutions. 
The neutron EDM measurement should be improved by more than two orders of magnitude, $d_n \lesssim 10^{-28}e~$cm~\cite{Chupp:2017rkp}.
There are also plans to measure the EDMs of charged nuclei such as the proton and deuteron in EM storage rings
with expected resolutions of $d_{p,D} \lesssim 10^{-29}e~$cm~\cite{Chupp:2017rkp}. Moreover, we expect also one order of magnitude improvement on 
the current measurement of molecular systems, such as the polar molecule ThO, which give rise to the most stringent constraints on the electron EDM 
and electron-nucleon couplings. If this is the case, the bounds on $d_G$ and quark (C)EDMs will improve by roughly three orders of magnitude while the bounds on the electron EDM and 4-fermion contributions will become one order of magnitude more stringent.

\bigskip
\section{Conclusions}
\label{sec:concl}
In this work, we have studied for the first time the full set of contributions of ALPs to permanent EDMs of molecules, atoms, nuclei and nucleons.
After classifying the CPV Jarlskog invariants emerging in the ALP EFT, we have evaluated the leading short-distance effects to EDMs up to two-loop order. 
Our main result is that 4-fermion and Weinberg operators, so far neglected in the literature, provide by far the largest contributions to the EDMs.

Our work can be generalised in several directions. 
For instance, it would be interesting to extend our analysis to ALP masses in the sub-GeV region where QCD cannot be treated perturbatively. 
Moreover, it could be worth to investigate possible UV completions of our EFT that resemble the strong dynamics of the pion in the SM. 
Furthermore, since relaxion models addressing the hierarchy problem share many similarities with our EFT, it would be interesting to check 
whether these scenarios can be probed by means of the new EDM observables studied here. 
Finally, another ambitious project would be to investigate whether a successful baryogenesis can be driven by CPV ALP interactions. 

In summary, a CPV ALP can be related to many fundamental open questions in particle physics. It is very exciting that the outstanding experimental 
progress, which is expected in the next years, on the searches for permanent EDMs will likely shed light on some of them.

\section*{Acknowledgments} 

We thank Stefano Bertolini, Martin Jung and Fabrizio Nesti for useful discussions. 
The work of LDL is supported by the Marie Sk\l{}odowska-Curie 
Individual Fellowship grant AXIONRUSH (GA 840791) and the Deutsche Forschungsgemeinschaft under Germany's Excellence Strategy - EXC 2121 Quantum Universe - 390833306.


\begin{thebibliography}{0}%
\makeatletter
\providecommand \@ifxundefined [1]{%
 \@ifx{#1\undefined}
}%
\providecommand \@ifnum [1]{%
 \ifnum #1\expandafter \@firstoftwo
 \else \expandafter \@secondoftwo
 \fi
}%
\providecommand \@ifx [1]{%
 \ifx #1\expandafter \@firstoftwo
 \else \expandafter \@secondoftwo
 \fi
}%
\providecommand \natexlab [1]{#1}%
\providecommand \enquote  [1]{``#1''}%
\providecommand \bibnamefont  [1]{#1}%
\providecommand \bibfnamefont [1]{#1}%
\providecommand \citenamefont [1]{#1}%
\providecommand \href@noop [0]{\@secondoftwo}%
\providecommand \href [0]{\begingroup \@sanitize@url \@href}%
\providecommand \@href[1]{\@@startlink{#1}\@@href}%
\providecommand \@@href[1]{\endgroup#1\@@endlink}%
\providecommand \@sanitize@url [0]{\catcode `\\12\catcode `\$12\catcode
  `\&12\catcode `\#12\catcode `\^12\catcode `\_12\catcode `\%12\relax}%
\providecommand \@@startlink[1]{}%
\providecommand \@@endlink[0]{}%
\providecommand \url  [0]{\begingroup\@sanitize@url \@url }%
\providecommand \@url [1]{\endgroup\@href {#1}{\urlprefix }}%
\providecommand \urlprefix  [0]{URL }%
\providecommand \Eprint [0]{\href }%
\providecommand \doibase [0]{http://dx.doi.org/}%
\providecommand \selectlanguage [0]{\@gobble}%
\providecommand \bibinfo  [0]{\@secondoftwo}%
\providecommand \bibfield  [0]{\@secondoftwo}%
\providecommand \translation [1]{[#1]}%
\providecommand \BibitemOpen [0]{}%
\providecommand \bibitemStop [0]{}%
\providecommand \bibitemNoStop [0]{.\EOS\space}%
\providecommand \EOS [0]{\spacefactor3000\relax}%
\providecommand \BibitemShut  [1]{\csname bibitem#1\endcsname}%
\let\auto@bib@innerbib\@empty
\end{thebibliography}%


\begin{thebibliography}{99}



\bibitem{Jaeckel:2010ni} 
  J.~Jaeckel and A.~Ringwald,
  Ann.\ Rev.\ Nucl.\ Part.\ Sci.\  {\bf 60}, 405 (2010);
%
D.~J.~E.~Marsh,
Phys. Rept. \textbf{643}, 1-79 (2016);
%
  I.~G.~Irastorza and J.~Redondo,
  Prog.\ Part.\ Nucl.\ Phys.\  {\bf 102} (2018) 89;
L.~Di Luzio, M.~Giannotti, E.~Nardi and L.~Visinelli,
Phys. Rept. \textbf{870}, 1-117 (2020).

\bibitem{Peccei:1977hh} 
  R.~D.~Peccei and H.~R.~Quinn,
  Phys.\ Rev.\ Lett.\  {\bf 38}, 1440 (1977);
%
Phys. Rev. D \textbf{16}, 1791-1797 (1977);
%
  S.~Weinberg,
  Phys.\ Rev.\ Lett.\  {\bf 40}, 223 (1978);
%
  F.~Wilczek,
  Phys.\ Rev.\ Lett.\  {\bf 40}, 279 (1978).

\bibitem{Preskill:1982cy} 
  J.~Preskill, M.~B.~Wise and F.~Wilczek,
  Phys.\ Lett.\  {\bf 120B}, 127 (1983);
%
  L.~F.~Abbott and P.~Sikivie,
  Phys.\ Lett.\  {\bf 120B}, 133 (1983);
%
  M.~Dine and W.~Fischler,
  Phys.\ Lett.\  {\bf 120B}, 137 (1983);
%
  R.~L.~Davis,
  Phys.\ Lett.\ B {\bf 180}, 225 (1986).

\bibitem{Wilczek:1982rv} 
  A.~Davidson and K.~C.~Wali,
  Phys.\ Rev.\ Lett.\  {\bf 48} (1982) 11;
  F.~Wilczek,
  Phys.\ Rev.\ Lett.\  {\bf 49}, 1549 (1982);
%
  Y.~Ema, K.~Hamaguchi, T.~Moroi and K.~Nakayama,
  JHEP {\bf 1701} (2017) 096;
%
  L.~Calibbi, F.~Goertz, D.~Redigolo, R.~Ziegler and J.~Zupan,
  Phys.\ Rev.\ D {\bf 95}, no. 9, 095009 (2017).

\bibitem{Graham:2015cka} 
  P.~W.~Graham, D.~E.~Kaplan and S.~Rajendran,
  Phys.\ Rev.\ Lett.\  {\bf 115}, no. 22, 221801 (2015).
    
\bibitem{Chang:2000ii}
  D.~Chang, W.~F.~Chang, C.~H.~Chou and W.~Y.~Keung,
  Phys.\ Rev.\ D {\bf 63} (2001) 091301.
    
\bibitem{Marciano:2016yhf}
  W.~J.~Marciano, A.~Masiero, P.~Paradisi and M.~Passera,
  Phys.\ Rev.\ D {\bf 94} (2016) no.11,  115033.
  
\bibitem{Krasznahorkay:2015iga}
  A.~J.~Krasznahorkay {\it et al.},
  Phys.\ Rev.\ Lett.\  {\bf 116} (2016) no.4,  042501.

\bibitem{Feng:2016ysn}
  J.~L.~Feng, B.~Fornal, I.~Galon, S.~Gardner, J.~Smolinsky, T.~M.~P.~Tait and P.~Tanedo,
  Phys.\ Rev.\ D {\bf 95} (2017) no.3,  035017.

\bibitem{Ellwanger:2016wfe}
  U.~Ellwanger and S.~Moretti,
  JHEP {\bf 1611} (2016) 039.

\bibitem{Aprile:2020tmw}
  E.~Aprile {\it et al.} [XENON Collaboration],
  Phys.\ Rev.\ D {\bf 102}, no.7,  072004, (2020).

\bibitem{Takahashi:2020bpq}
F.~Takahashi, M.~Yamada and W.~Yin,
Phys. Rev. Lett. \textbf{125}, no.16, 161801 (2020);
  I.~M.~Bloch, A.~Caputo, R.~Essig, D.~Redigolo, M.~Sholapurkar and T.~Volansky,
  arXiv:2006.14521 [hep-ph].

\bibitem{Georgi:1986df} 
  H.~Georgi, D.~B.~Kaplan and L.~Randall,
  Phys.\ Lett.\  {\bf 169B}, 73 (1986).

\bibitem{Dobrich:2019dxc}
B.~D\"obrich, J.~Jaeckel and T.~Spadaro,
JHEP \textbf{05}, 213 (2019)
[erratum: JHEP \textbf{10}, 046 (2020)].

\bibitem{Jaeckel:2015jla} 
  J.~Jaeckel and M.~Spannowsky,
  Phys.\ Lett.\ B {\bf 753}, 482 (2016);
%
  S.~Knapen, T.~Lin, H.~K.~Lou and T.~Melia,
  Phys.\ Rev.\ Lett.\  {\bf 118}, no. 17, 171801 (2017);
%
  I.~Brivio, {\it et al.},
  Eur.\ Phys.\ J.\ C {\bf 77}, no. 8, 572 (2017);
%
  A.~Mariotti, D.~Redigolo, F.~Sala and K.~Tobioka,
  Phys.\ Lett.\ B {\bf 783} (2018) 13;
%
  X.~Cid Vidal, A.~Mariotti, D.~Redigolo, F.~Sala and K.~Tobioka,
  JHEP {\bf 1901} (2019) 113
   Erratum: [JHEP {\bf 2006} (2020) 141];
%
  D.~Aloni, Y.~Soreq and M.~Williams,
  Phys.\ Rev.\ Lett.\  {\bf 123} (2019) no.3,  031803;
%
  D.~Aloni, C.~Fanelli, Y.~Soreq and M.~Williams,
  Phys.\ Rev.\ Lett.\  {\bf 123}, no. 7, 071801 (2019).
%

\bibitem{Bauer:2017ris} 
  M.~Bauer, M.~Neubert and A.~Thamm,
  JHEP {\bf 1712}, 044 (2017);
%
  Phys.\ Rev.\ Lett.\  {\bf 119} (2017) no.3,  031802.

\bibitem{Batell:2009jf}
  B.~Batell, M.~Pospelov and A.~Ritz,
  Phys.\ Rev.\ D {\bf 83} (2011) 054005;
%
  M.~B.~Gavela, R.~Houtz, P.~Quilez, R.~Del Rey and O.~Sumensari,
  Eur.\ Phys.\ J.\ C {\bf 79} (2019) no.5,  369;
%
J.~Martin Camalich, M.~Pospelov, P.~N.~H.~Vuong, R.~Ziegler and J.~Zupan,
Phys. Rev. D \textbf{102}, no.1, 015023 (2020).
    
\bibitem{Bauer:2019gfk}
  M.~Bauer, M.~Neubert, S.~Renner, M.~Schnubel and A.~Thamm,
  Phys.\ Rev.\ Lett.\  {\bf 124} (2020) no.21,  211803;
%
  C.~Cornella, P.~Paradisi and O.~Sumensari,
  JHEP {\bf 2001} (2020) 158;
%
  L.~Calibbi, D.~Redigolo, R.~Ziegler and J.~Zupan,
  arXiv:2006.04795 [hep-ph].

  
\bibitem{Moody:1984ba}
J.~E.~Moody and F.~Wilczek,
Phys. Rev. D \textbf{30}, 130 (1984);
%
  M.~Pospelov,
  Phys.\ Rev.\ D {\bf 58} (1998) 097703;
%
G.~Raffelt,
Phys. Rev. D \textbf{86}, 015001 (2012);
S.~Bertolini, L.~Di Luzio and F.~Nesti,
  Phys.\ Rev.\ Lett.\  {\bf 126} (2021) no.8,  081801;
C.~A.~J.~O'Hare and E.~Vitagliano,
Phys.\ Rev.\ D {\bf 102} (2020) no.11, 115026.

\bibitem{Stadnik:2017hpa}
Y.~V.~Stadnik, V.~A.~Dzuba and V.~V.~Flambaum,
Phys. Rev. Lett. \textbf{120}, no.1, 013202 (2018);
V.~A.~Dzuba, V.~V.~Flambaum, I.~B.~Samsonov and Y.~V.~Stadnik,
Phys. Rev. D \textbf{98}, no.3, 035048 (2018). 

\bibitem{Choi:1990cn}
  K.~Choi and J.~y.~Hong,
  Phys.\ Lett.\ B {\bf 259} (1991) 340.

\bibitem{Chupp:2017rkp}
T.~Chupp, P.~Fierlinger, M.~Ramsey-Musolf and J.~Singh,
Rev. Mod. Phys. \textbf{91}, no.1, 015001 (2019).
%
\bibitem{paper_long} L.~Di~Luzio, R.~Gr\"{o}ber, and P.~Paradisi, work in progress. 
%


\bibitem{Shifman:1978zn}
M.~A.~Shifman, A.~I.~Vainshtein and V.~I.~Zakharov,
Phys. Lett. B \textbf{78}, 443-446 (1978).
%
\bibitem{Jarlskog:1985ht}
  C.~Jarlskog,
  Phys.\ Rev.\ Lett.\  {\bf 55} (1985) 1039.
%
\bibitem{Pospelov:2005pr}
  M.~Pospelov and A.~Ritz,
  Annals Phys.\  {\bf 318} (2005) 119;
  %
M.~Pospelov and A.~Ritz,
Phys. Rev. D \textbf{63}, 073015 (2001); 
O.~Lebedev, K.~A.~Olive, M.~Pospelov and A.~Ritz,
Phys. Rev. D \textbf{70}, 016003 (2004);
J.~R.~Ellis, J.~S.~Lee and A.~Pilaftsis,
JHEP \textbf{10}, 049 (2008).

\bibitem{Chen:2015vqy}
  C.~Y.~Chen, H.~Davoudiasl, W.~J.~Marciano and C.~Zhang,
  Phys.\ Rev.\ D {\bf 93} (2016) no.3,  035006.

\bibitem{Degrassi:2005zd}
  G.~Degrassi, E.~Franco, S.~Marchetti and L.~Silvestrini,
  JHEP {\bf 0511} (2005) 044;
%
  J.~Hisano, K.~Tsumura and M.~J.~S.~Yang,
  Phys.\ Lett.\ B {\bf 713} (2012) 473;
%
  E.~E.~Jenkins, A.~V.~Manohar and P.~Stoffer,
  JHEP {\bf 1801} (2018) 084.



\bibitem{Dekens:2018bci}
  W.~Dekens, J.~de Vries, M.~Jung and K.~K.~Vos,
  JHEP {\bf 1901} (2019) 069.
%

\bibitem{Andreev:2018ayy}
  V.~Andreev {\it et al.} [ACME Collaboration],
  Nature {\bf 562} (2018) no.7727,  355;
  J.~Baron {\it et al.} [ACME Collaboration],
  Science {\bf 343} (2014) 269.

\bibitem{Hisano:2012cc}
J.~Hisano, K.~Tsumura and M.~J.~S.~Yang,
Phys. Lett. B \textbf{713}, 473-480 (2012).

\bibitem{Cirigliano:2019vfc}
  V.~Cirigliano, A.~Crivellin, W.~Dekens, J.~de Vries, M.~Hoferichter and E.~Mereghetti,
  Phys.\ Rev.\ Lett.\  {\bf 123} (2019) no.5,  051801.

\bibitem{Bertolini:2019out}
S.~Bertolini, A.~Maiezza and F.~Nesti,
Phys. Rev. D \textbf{101}, no.3, 035036 (2020).

\bibitem{Abel:2020gbr}
C.~Abel \textit{et al.} [nEDM],
Phys. Rev. Lett. \textbf{124}, no.8, 081803 (2020);
  J.~M.~Pendlebury {\it et al.},
  Phys.\ Rev.\ D {\bf 92} (2015) no.9,  092003.

\bibitem{Graner:2016ses}
  B.~Graner, Y.~Chen, E.~G.~Lindahl and B.~R.~Heckel,
  Phys.\ Rev.\ Lett.\  {\bf 116} (2016) no.16,  161601
   Erratum: [Phys.\ Rev.\ Lett.\  {\bf 119} (2017) no.11,  119901];
  W.~C.~Griffith, M.~D.~Swallows, T.~H.~Loftus, M.~V.~Romalis, B.~R.~Heckel and E.~N.~Fortson,
  Phys.\ Rev.\ Lett.\  {\bf 102} (2009) 101601.

\bibitem{Giudice:2012ms}
G.~F.~Giudice, P.~Paradisi and M.~Passera,
JHEP \textbf{11} (2012), 113.






\end{thebibliography}
\end{document}